\documentstyle[epsfig]{aipproc}

\def \gray     {$\gamma$-ray}
\def \grays    {$\gamma$-rays}

\def \sig      {$\sigma$}

\begin{document}
\title{Multifrequency Observations of the Virgo Blazars 3C 273 and 3C 279 
       in CGRO Cycle 8}

\author{W. Collmar$^1$, S. Benlloch$^2$, J.E. Grove$^3$, R.C. Hartman$^4$,\\
        W.A. Heindl$^5$, A. Kraus$^6$, H. Ter\"asranta$^7$, M. Villata$^8$,
        K. Bennett$^9$,\\
        H. Bloemen$^{10}$, W.N. Johnson$^3$, T.P. Krichbaum$^6$,
        C.M. Raiteri$^8$,\\
        J. Ryan$^{11}$, G. Sobrito$^8$, V. Sch\"onfelder$^1$,
        O.R. Williams$^9$, J. Wilms$^2$
       }
\address{
$^1$Max-Planck-Institut f\"ur extraterrestrische Physik, Postfach1603, 85740 Garching, Germany\\
$^2$Institut f\"ur Astronomie und Astrophysik, Univ. of T\"ubingen, T\"ubingen, Germany\\
$^3$Naval Research Lab., 4555 Overlook Av., SW, Washington, DC 20375-5352, USA\\
$^4$NASA/Goddard Space Flight Center, Greenbelt, MD 20771\\
$^5$Center for Astrophysics and Space Sciences, UCSD, La Jolla, CA, USA\\
$^6$Max-Planck-Institut f\"ur Radioastronomie, D-53121 Bonn, Germany\\
$^7$Mets\"ahovi Radio Research Station, FIN-02540 Kylm\"al\"a, Finland\\
$^8$Osservatorio Astronomico di Torino, I-10025 Pino Torinese, Italy\\
$^9$Astrophysics Division, ESTEC, NL-2200 AG Noordwijk, The Netherlands\\
$^{10}$SRON-Utrecht, Sorbonnelaan 2, NL-3584 CA Utrecht, The Netherlands\\
$^{11}$Universtity of New Hampshire, Durham NH 03824-3525, USA\\
  }

\maketitle

\begin{abstract}
We report first observational results of multifrequency campaigns on the 
prominent Virgo blazars 3C~273 and 3C~279 which were carried out in January 
and February 1999. Both blazars are detected from radio to \gray\ energies. 
We present the measured X- to \gray\ spectra of both sources, and for 
3C~279 we compare the 1999 broad-band (radio to \gray) spectrum to 
measured previous ones.
\end{abstract}

\section*{Introduction}
We report on simultaneous multifrequency observations of the prominent Virgo 
blazars 3C~273 and 3C~279 during CGRO Cycle 8. Because both blazars are known \gray\ sources, which have been detected by the CGRO experiments several 
times before, we proposed for simultaneous CGRO (OSSE, COMPTEL) and 
RXTE high-energy observations. The prime goal was to simultaneously 
measure their high-energy spectra from about 2.5~keV to 30~MeV. Because of 
the shortage of spark chamber gas, the EGRET experiment is hardly available 
anymore and therefore was not requested in the proposals. 
After the proposed simultaneous high-energy observations were approved
and scheduled, additional simultaneous observations were performed from ground-based observers extending the energy range of the campaigns to
lower energies.

In this paper we report first observational results of the campaigns 
with emphasis on the X- and \gray\ part. In particular we present the measured X- to \gray\ spectra of both sources.     

\section*{Observations}
The multifrequency observations were carried out between 1999 January 5 and February 2. The observational strategy was that both blazars are within the COMPTEL field-of-view for the whole 4 weeks reaching the optimal sensitivity for the \gray\ observations, and OSSE observes simultaneously each source for 2 weeks. Within these two-week OSSE periods, three RXTE pointings were scheduled for each source covering simultaneously the X- and hard X-ray part of the spectrum and providing information on the X-ray variability.  
To supplement these high-energy observations both sources were simultaneously  observed in different optical and radio bands. On 1999 January 15 the optical flux of 3C~279 reached a high level which -- after some discussion -- triggered  target-of-opportunity (ToO) observations of this source. This resulted in a switch-on of EGRET which led to EGRET observations of both quasars during roughly the second half of the campaigns. Because of the 3C~279 ToO, the OSSE 3C~273 observation was stopped after one week, and OSSE observed 3C~279 again. These EGRET and OSSE ToO observations of 3C~279 are property of a different
CGRO proposal and therefore their results are not reported here. The detailed high-energy observational log is given in Table~\ref{tab1}.

\begin{table}[t!]
\caption{
Summary of the satellite observations of both Virgo blazars during the
campaigns in 1999. The observation periods as well as the coverage in energy
are given.  
}
\label{tab1}
\begin{tabular}{lccc}
 Source      & Experiment & Obs. Period in '99 & Energy Band \\
\tableline
3C~273       & EGRET    & Jan. 20 - Feb. 2  & 30 MeV - 10 GeV        \\
             & COMPTEL  & Jan. 5 - Feb. 2   & 750 keV - 30 MeV       \\
             & OSSE     & Jan. 19 - Jan. 26 &  50 keV - $\sim$1 MeV  \\
             & RXTE \tablenote{3 individual pointings of $\sim$13~ksec each} 
& Jan 19, 26, Feb. 1 & 2.5~keV - $\sim$200 keV  \\
\tableline
3C~279       & EGRET\tablenote{3C~279 ToO observation; not reported here}    & Jan. 20 - Feb. 2  & 30 MeV - 10 GeV        \\
             & COMPTEL  & Jan. 5 - Feb. 2   & 750 keV - 30 MeV       \\
             & OSSE     & Jan. 5 - Jan. 19  &  50 keV - $\sim$1 MeV  \\
             & OSSE$^{b}$ & Jan. 26 - Feb. 2  &  50 keV - $\sim$1 MeV  \\
             & RXTE$^{a}$ 
& Jan 5, 12, 17 & 2.5~keV - $\sim$200 keV  \\
\end{tabular}\end{table}

\section*{Results}
\subsection*{3C~273}

\begin{figure}[t!] 
\centerline{\epsfig{file=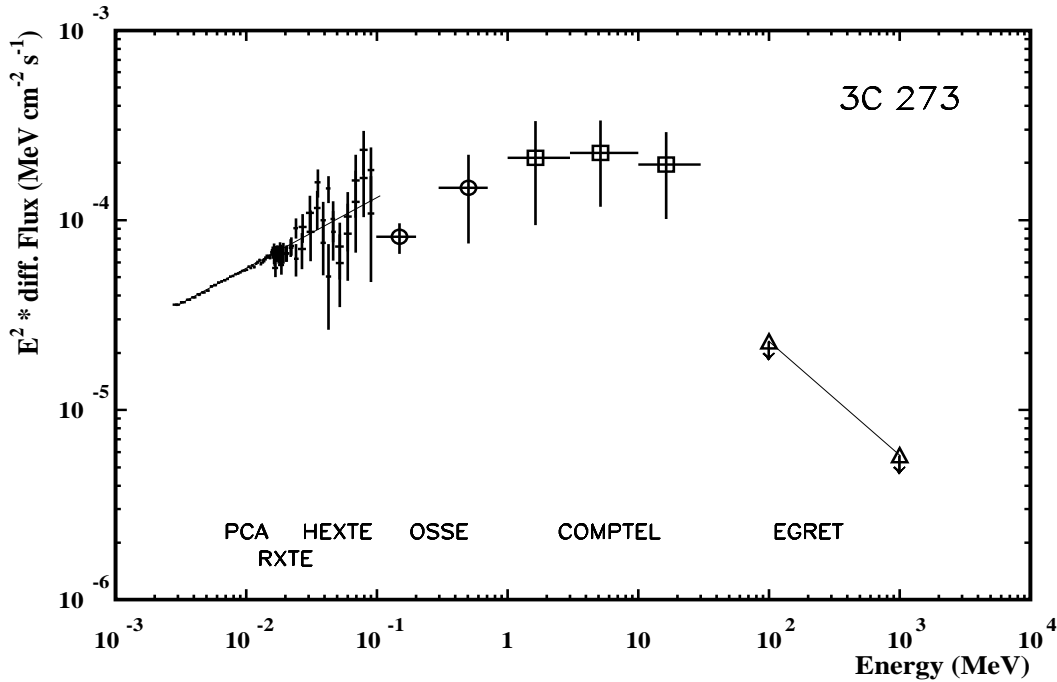,height=4.0in,width=6.0in,angle=0.,clip=}}
\vspace{0.0cm}
\caption{Simultaneous X-ray and \gray\ spectrum of 3C~273 as observed during 
the January/February 1999 campaign. The data points from the different experiments are derived from their total observation times during the campaign, which are given in Table~\ref{tab1}. The RXTE (+) spectral points are derived 
from the observation sum of the three individual pointings, and are shown together with the best-fit power-law spectrum (solid line). EGRET did not
detect the quasar at energies above 100~MeV providing an upper limit on the flux. For the upper limit line, drawn between 100~MeV and 1~GeV, a spectral
power-law shape with photon index of 2.6 is assumed. The error bars are 1\sig. 
}
\label{fig1}
\end{figure}

3C~273 is significantly detected in all observed low-energy (radio, optical) 
bands. At high energies the quasar is significantly detected in X- and hard 
X-rays by RXTE from about 2.5 to $\sim$100~keV, showing a power-law spectrum with a photon index (E$^{-\alpha}$) $\alpha$ of 1.6. OSSE detects the source in the one-week observation at a significance level of $\sim$5\sig.
Therefore its spectrum had to be rebinned severely to reach two significant spectral points. COMPTEL detects the 
blazar in the sum of the 4-week observation at the $\sim$5\sig\ level. However, 
despite the COMPTEL detection at MeV-energies, EGRET -- covering only half 
of the COMPTEL observation time -- does not detect 3C~273 at energies above 
100~MeV. The combined -- more or less simultaneous -- high-energy spectrum 
of 3C~273 is shown in Figure~\ref{fig1}. The well-known bending 
(e.g. \cite{Lichti95}, \cite{Montigny97}) at MeV-energies is visible. 
The most surprising result however, is the 
non-detection by EGRET at high-energy \grays, despite the COMPTEL 
detection at MeV-energies. This requires a strong spectral turnover between 
30 and 100~MeV, and might hint at different generation mechanisms
for the MeV and $>$100~MeV photon populations.

\subsection*{3C~279}
\begin{figure}[t!]
\centerline{\epsfig{file=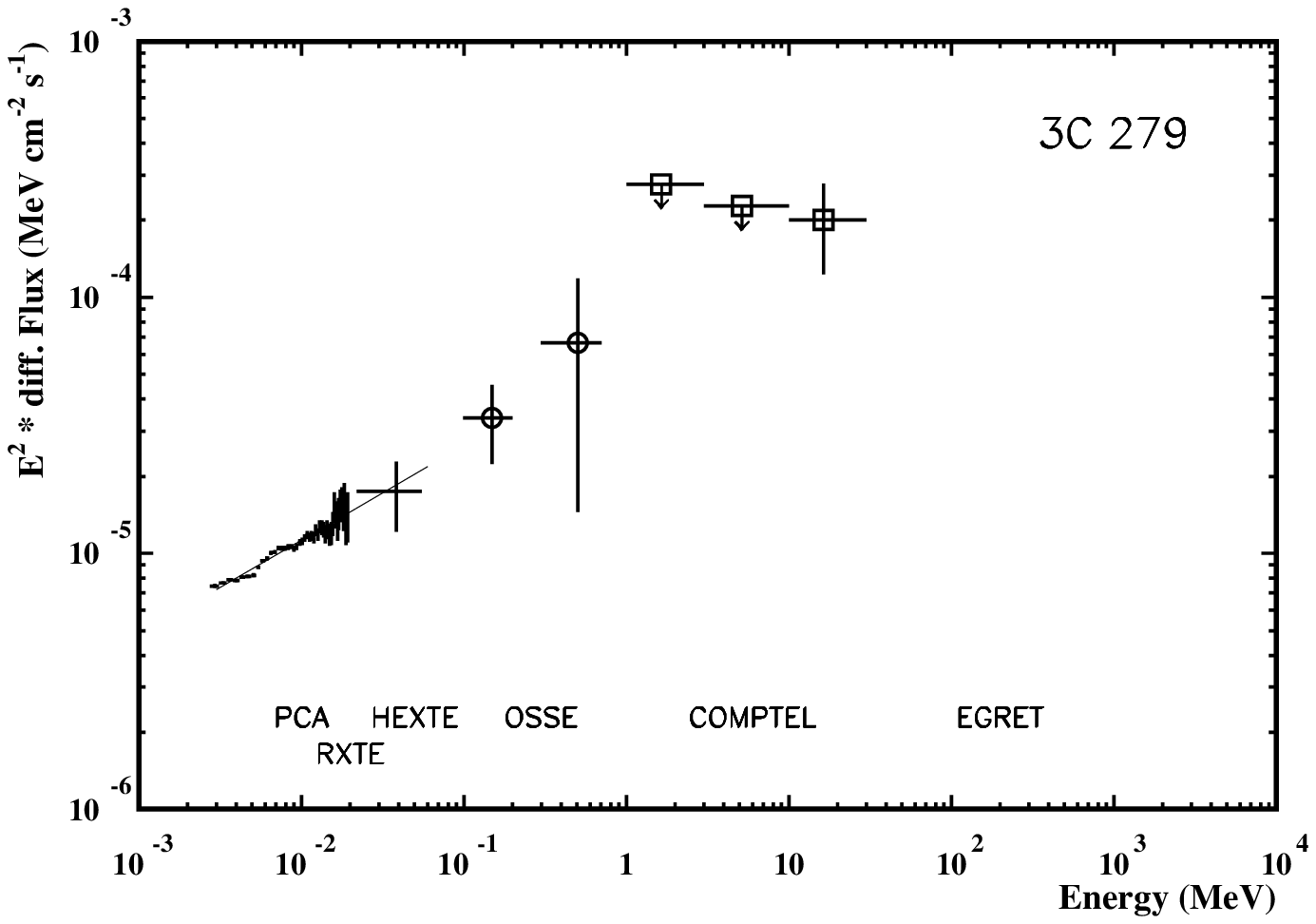,height=4.0in,width=6.0in,angle=0.,clip=}}
\vspace{0.0cm}
\caption{Quasi-simultaneous keV- to MeV-spectrum of the \gray\ blazar 3C~279 as observed in early 1999. The data points from the different experiments are derived from their total observations times during the campaign, which are given in Table~\ref{tab1}. The RXTE (+) spectral points are derived 
from the observation sum of the three individual pointings, and are shown together with the best-fit power-law spectrum (solid line). The error bars are 1\sig\ and the upper limits are 2\sig. Simultaneous EGRET observations exist 
(Table~\ref{tab1}), which will be reported elsewhere. The spectrum can be
described by a single power-law shape from about 2.5~keV to 30~MeV.}
\label{fig2}
\end{figure}

\begin{figure}[t!]
\centerline{\epsfig{file=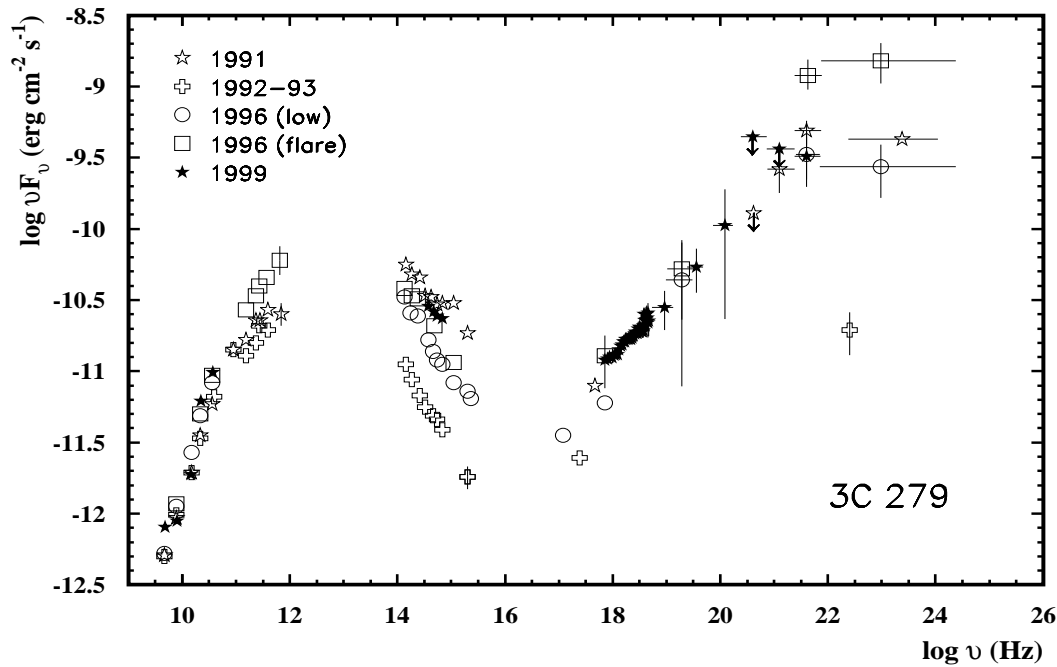,height=4.0in,width=6.0in,angle=0.,clip=}}
\vspace{0.0cm}
\caption{Broad-band spectrum of 3C~279 for different epochs. The 1999 results (filled stars) reported here fit nicely to the two previous measurements in 1991 and 1996 when 3C~279 was in a \gray\ high state. The spectral points of the earlier measurements are taken from [4].} 
\label{fig3}
\end{figure}

3C~279 is significantly detected at the radio and optical bands, showing 
strong time variability and flaring activity in the optical. 
At X-rays the blazar is significantly observed up to 20~keV by the RXTE/PCA, 
and is detected in hard X-rays between 20 and 50~keV by RXTE/HEXTE with a significance of about 5\sig.
The RXTE/PCA spectrum is well fitted by a single power-law model 
with an index $\alpha$ of 1.6. At higher energies the detection significances 
become marginal. OSSE found 3\sig-evidence for the source only at their lower 
energies (near 100~keV), and COMPTEL -- also at the 3\sig\ level -- only at 
their upper energies. At energies above 100~MeV EGRET has significantly
detected 3C~279 ([3]), however these data are not reported here. 
The measured X- to \gray\ spectrum of 3C~279 is given in
Figure~\ref{fig2}. It shows that 3C~279 was observed in a bright \gray\ state. The flux at the COMPTEL upper energies is at the same level as measured during the two previous \gray\ high states in 1991 and 1996 (Figure~\ref{fig3}).
The spectral power-law shape measured from $\sim$2.5 to 20~keV can -- according to the current state of analysis -- be extrapolated up to 30~MeV without any obvious breaks or bendings. This suggests that this part of the spectrum, which is considered to be non-thermal inverse-Compton radiation, is emitted
by a single emission component or mechanism.     
  
\begin{references}
\bibitem{Lichti95}Lichti, G.G., et al., {\it A\&A} {\bf 298}, 711 (1995).
\bibitem{Montigny97}v. Montigny, C., et al., {\it ApJ} {\bf 483}, 161 (1997).
\bibitem{Hartman99}Hartman, R.C., {\it priv. comm.}, (1999).
\bibitem{Wehrle98}Wehrle, A.E., et al., {\it ApJ} {\bf 497}, 178 (1998).
\end{references}
\end{document}